\documentclass[10pt]{article}
\usepackage[dvips]{epsfig}
\usepackage{epsf}
\newcommand{\beq}{\begin{equation}}

\newcommand{\eeq}{\end{equation}}
\newcommand{\bea}{\begin{eqnarray}}
\newcommand{\eea}{\end{eqnarray}}
\def\barlambda{{\lambda\kern -0.5em \raise 0.5 ex \hbox{--}}}
\begin{document}

\begin{center}

{\large \bf Extension of the LAQGSM03.01 Code to Describe Photo-Nuclear
Reactions \\ up to Tens of GeV
}\\

\vspace{0.5cm}

K.~K.~Gudima$^1$ and S.~G.~Mashnik$^2$

\vspace{0.2cm}

$^1${\em Institute of Applied Physics, Academy of Science of Moldova,
Chi\c{s}in\u{a}u, Moldova}\\

$^2${\em  
X-3-MCC, Los Alamos National Laboratory, Los Alamos, NM 87545, USA}\\

\end{center}

\vspace{0.3cm}
\begin{center}
{\bf Abstract}
\end{center}
{\noindent
The Los Alamos version of the Quark-Gluon 
String Model implemented in the code
LAQGSM03 was  extended 
to describe photonuclear reactions at energies up to tens of GeV.
We have incorporated 56 channels to consider  
$\gamma p$ elementary interactions during the cascade stage of
reactions and 56 channels for  $\gamma n$ interactions,
in addition to absorption of photons on two-nucleon pairs,
 into LAQGSM03
and have tested the extended code referred to as LAQGSM03.01
on a number of measured high-energy photonuclear reactions.
Here, we present the photonuclear reaction model of
LAQGSM03.01 and show examples of its results compared
with available experimental data.
}

\vspace*{5mm}
{\noindent \bf \large 1. Introduction}\\

Following an increased interest in 
nuclear data 
for such projects as the Accelerator Transmutation of nuclear 
Wastes (ATW), 
Accelerator Production of Tritium (APT), 
Spallation Neutron Source (SNS), 
Rare Isotope Accelerator (RIA), 
Proton Radiography (PRAD) as a radiographic probe for the Advanced 
Hydro-test Facility and others, for several years the US Department 
of Energy has supported our work on the development of 
improved versions of the Cascade-Exciton Model (CEM)
of nuclear reactions \cite{JNRS05,CEM03.01}
and the Los Alamos version of the Quark-Gluon String Model (LAQGSM) 
\cite{LAQGSM,Varenna03}
to describe reactions induced by particles and
nuclei at energies up to about 1 TeV/nucleon \cite{TRAMU}-\cite{ResNote06}.
To describe fission and production of light fragments heavier
than $^4$He, we have
merged both our codes with several evaporation/fission/fragmentation models,
including the Generalized Evaporation/fission Model code
GEM2 by Furihata \cite{GEM2},
the fission-like binary-decay code GEMINI by Charity 
{\it et al.} \cite{GEMINI},
and the Statistical Multifragmentation Model (SMM) by Botvina
{\it et al.} \cite{SMM}.
Our codes
perform as well as and often better than other current 
models in describing a large variety of spallation,
fission, and fragmentation reactions, therefore they are used
as event-generators in several transport codes.

However, the initial version of the Los Alamos Quark-Gluon String 
Model (LAQGSM) \cite{LAQGSM}, just like its precursor, the Quark-Gluon 
String Model (QGSM) \cite{QGSM}, did not consider photonuclear
reactions, while its 2003 version, LAQGSM03 \cite{JNRS05}, describes
such reactions only for energies up to about 1.5 GeV.
This is not convenient when using our codes to solve 
problems for PRAD, NASA, and other high-energy applications 
or to analyze future high-energy measurements at the
Thomas Jefferson National Accelerator Facility (CEBAF),
where photons with much higher energy are created and 
need to be simulated with our even-generators
in transport codes like MCNP6 \cite{MCNP6}, MARS \cite{MARS}, 
MCNPX \cite{MCNPX}, and others. 
To address this problem, we have extended \cite{ResNote05}
LAQGSM03.01
to describe photonuclear reactions up to tens of GeV.
The next section presents a short description of the high-energy
photonuclear reaction model of LAQGSM03.01, followed by two
sections with examples of calculated
particle spectra, yields, and recoil properties of products
from high-energy photonuclear reactions compared with
available data.\\

{\noindent \bf \large 2. High-Energy Photonuclear Model of LAQGSM03.01}\\

Initially \cite{JNRS05}, an improved version of the
Dubna InraNuclear cascade photonuclear reaction model 
developed originally 37 years ago by one of us in collaboration  
 with Iljinov and Toneev \cite{Dubna69}
to describe photonuclear reactions at energies
above the Giant Dipole Resonance (GDR) region was incorporated
into LAQGSM.
[At photon energies $T_{\gamma} = 10$--$40$ MeV, the de Broglie 
wavelength $\barlambda$ is of the order of $20$--$5$ fm,  
greater than the average inter-nucleonic distance in the nucleus; 
the photons interact with the nuclear
dipole resonance as a whole, thus the INC is not applicable.]
Below the pion production threshold, the Dubna INC considers
absorption of photons on ``quasi-deuteron" pairs according
to the Levinger model \cite{Levinger}:
\begin{equation}
\sigma_{\gamma A} = L \frac{Z(A-Z)}{A} \sigma_{\gamma d} \mbox{ ,} 
\end{equation}
where $A$ and $Z$ are the mass and charge numbers of the nucleus,
$L \approx 10$, and $ \sigma_{\gamma d}$ is 
the total photoabsorption cross section on deuterons as
defined from experimental data.

At photon energies above the pion-production threshold, the Dubna INC
considers production of only
one or two pions (channels \#1--3 and 15--17 from Table 1); the concrete
mode of the reaction is chosen by the Monte Carlo method according to the
partial cross sections, defined from available experimental data. This
limits the use of such an approach to photon energies of only up to about
1.5 GeV, as contribution from multiple pion production
become predominant at higher energies. 

To address this problem,
we have extended LAQGSM03 \cite{JNRS05} to describe photonuclear reactions
at energies up to tens of GeV. For this, we took advantage of
the high-energy event generators for $\gamma p$ and  $\gamma n$
elementary interactions from the Moscow high-energy
photonuclear reaction model \cite{Iljinov97} kindly sent us
by one of its coauthors, Dr. Igor Pshenichnov.
Let us note that the $\gamma p$ and  $\gamma n$ event generators
from the  Moscow INC \cite{Iljinov97} use an improvement and
extension to higher energies of the Genova 
$\gamma p$ and  $\gamma n$  event generators
developed by P. Corvisiero {\it et al.} to describe
photon-nucleon interactions up to 4 GeV \cite{GenovaEvGen}.
We have incorporated into LAQGSM03.01 56 channels to consider  
$\gamma p$ elementary interactions during the cascade stage of
reactions, and 56 channels for  $\gamma n$ interactions.
These reaction channels new to LAQGSM03.01 are listed in 
Table 1 (its precursor, LAQGSM03  \cite{JNRS05},
considers only channels \#1--3 and 15--17, and only
up to about 1.5 GeV). 

To describe in LAQGSM03.01 the two-body
channels  \#1--14, we use part of a file containing 
smooth approximations through presently
available experimental data sent us by Dr.\ Pshenichnov
but have developed our own algorithms and
written our own routines to simulate unambiguously
$d \sigma / d \Omega$ and to
choose the corresponding value of $\Theta$ for any $E_\gamma$,
using a single random number $\xi$ uniformly
distributed in the interval [0,1], as described in \cite{JNRS05}.
Fig. 1 shows examples of 
angular distributions of $\pi^{0}$ from
$\gamma p \to \pi^{0} p$ interactions 
as functions of $\Theta^{\pi}_{c.m.s}$ at eight different
photon energies as simulated by our routines compared with 
available experimental data. Examples of similar distributions
for  $\pi^{+}$ from
$\gamma p \to \pi^{+} n$ interactions may be found in \cite{JNRS05}.

To describe the channels \#15--21
with two and three pions in the final state, 
we use the $\gamma p$ and  $\gamma n$
event generators send us by Dr.\ Pshenichnov, but use our own
interpolation for integral cross sections. We do not show
here examples of cross sections for these channels, as we
reproduce with
LAQGSM03.01 all
the results by Pshenichnov {\it et al.}
shown in Figs. 6 and 8 of Ref. \cite{Iljinov97}.

Finally, to describe in LAQGSM03.01 the multi-pion channels
\#22--56, we use the isospin statistical model as
realized in the  $\gamma p$ and  $\gamma n$
event generators send us by Dr.\ Pshenichnov
and described in details in \cite{Iljinov97}, without any 
changes. 
For channels \#22--56,
we reproduce  exactly in LAQGSM03.01
the results by  Pshenichnov {\it et al.}
as published in Ref. \cite{Iljinov97}, therefore we do not show 
here examples of
such results.
 
After the bombarding photon is absorbed by two nucleons
or interacts inelastically with
a nucleon according the channels \#1--56, we get 
inside the nucleus several ``secondary" cascade
nucleons, pions, or other mesons and resonances listed in
Table 1, depending on which channel is simulated  from the  
corresponding cross sections at the given photon energy
to actually occur. These ``secondary" cascade particles interact further 
with inranuclear nucleons or leave the nucleus, depending
on their coordinates and momenta. The further behavior of the reaction
starting from this stage, after the photon had ``disappeared", is described 
by LAQGSM03.01 exactly the same way as for any other types of reactions,
induced, e.g., by nucleons or heavy ions.

LAQGSM03.01 \cite{ResNote05} is the latest modification of the  
Los Alamos version of the Quark-Gluon String Model \cite{LAQGSM},
which in its turn is an improvement of the  
Quark-Gluon String Model \cite{QGSM}.
It describes reactions induced by both particles and nuclei as
a three-stage process: IntraNuclear Cascade (INC), followed
by preequilibrium emission of particles during the equilibration of the
excited residual nuclei formed after the INC, followed by
evaporation of particles from or fission of the compound nuclei.
The INC stage of reactions is described 
with a recently improved version \cite{ResNote05} 
of the time-depending intranuclear cascade
model developed initially at Dubna, often 
referred in the literature simply
as the {\bf D}ubna intranuclear {\bf C}ascade {\bf M}odel, DCM
(see \cite{DCM} and references therein).
The preequilibrium part of reactions is described with
the last version 
of the Modified Exciton Model (MEM) from the improved
Cascade-Exciton Model (CEM) code CEM03.01 \cite{CEM03.01}. 
The evaporation and fission stages of reactions
are calculated with an updated and improved version of
the Generalized Evaporation Model code GEM2 by Furihata \cite{GEM2},
which considers evaporation of up to 66 types of different particles
and light fragments (up to $^{28}$Mg). If the excited residual
nucleus produced after the INC has a mass number $A < 13$,
LAQGSM03.01 uses a recently updated and improved
version of the Fermi Break-up model (in comparison with the initial
version developed in the group of Prof. V. S. Barashenkov at JINR, Dubna
and
used in QGSM  \cite{QGSM}

\clearpage

\vspace*{-10mm}
\begin{center}
Table 1. Channels of elementary $\gamma N$ interactions 
taken into account in LAQGSM03.01

\vspace{2mm}
\begin{small}
\begin{tabular}{r|l|l}
\hline\hline 
\# & $\gamma p$-interactions & $\gamma n$-interactions \\
\hline
1 & $\gamma p \rightarrow \pi^+ n$ &  $\gamma n  \rightarrow \pi^- p$ \\
2 & $\gamma p \rightarrow \pi^0 p$ &  $\gamma n  \rightarrow \pi^0 n$ \\
3 & $\gamma p \rightarrow \Delta^{++} \pi^-$ &
    $\gamma n \rightarrow \Delta^+    \pi^-$ \\
4 & $\gamma p \rightarrow \Delta^{+} \pi^0$ &
    $\gamma n \rightarrow \Delta^0   \pi^0$ \\
5 & $\gamma p \rightarrow \Delta^{0} \pi^+$ &
    $\gamma n \rightarrow \Delta^-   \pi^+$ \\
6 & $\gamma p \rightarrow \rho^0 p$ & $\gamma n \rightarrow \rho^0 n$ \\
7 & $\gamma p \rightarrow \rho^+ n$ & $\gamma n \rightarrow \rho^- p$ \\
8 & $\gamma p \rightarrow \eta p$ & $\gamma n \rightarrow \eta n$ \\
9 & $\gamma p \rightarrow \omega p$ & $\gamma n \rightarrow \omega n$ \\    
10 & $\gamma p \rightarrow \Lambda K^+$ & $\gamma n \rightarrow \Lambda K^0$\\ 
11 & $\gamma p \rightarrow \Sigma^0 K^+$ 
   & $\gamma n \rightarrow \Sigma^0 K^0$ \\
12 & $\gamma p \rightarrow \Sigma^+ K^0$ 
   & $\gamma n \rightarrow \Sigma^- K^+$ \\
13 & $\gamma p \rightarrow \eta' p$ & $\gamma n \rightarrow \eta' n$ \\
14 & $\gamma p \rightarrow \phi p$ & $\gamma n \rightarrow \phi n$ \\
\hline
15 & $\gamma p \rightarrow \pi^+ \pi^- p$ & 
     $\gamma n \rightarrow \pi^+ \pi^- n$ \\
16 & $\gamma p \rightarrow \pi^0 \pi^+ n$ & 
     $\gamma n \rightarrow \pi^0 \pi^- p$ \\
17 & $\gamma p \rightarrow \pi^0 \pi^0 p$ & 
     $\gamma n \rightarrow \pi^0 \pi^0 n$ \\
18 & $\gamma p \rightarrow \pi^0 \pi^0 \pi^0 p$ & 
     $\gamma n \rightarrow \pi^0 \pi^0 \pi^0 n$ \\
19 & $\gamma p \rightarrow \pi^+ \pi^- \pi^0 p$ & 
     $\gamma n \rightarrow \pi^+ \pi^- \pi^0 n$ \\
20 & $\gamma p \rightarrow \pi^+ \pi^0 \pi^0 n$ & 
     $\gamma n \rightarrow \pi^- \pi^0 \pi^0 p$ \\
21 & $\gamma p \rightarrow \pi^+ \pi^+ \pi^- n$ & 
     $\gamma n \rightarrow \pi^+ \pi^- \pi^- p$ \\
\hline
22 & $\gamma p \rightarrow \pi^0 \pi^0 \pi^0 \pi^0 p$ & 
     $\gamma n \rightarrow \pi^0 \pi^0 \pi^0 \pi^0 n$ \\
23 & $\gamma p \rightarrow \pi^+ \pi^- \pi^0 \pi^0 p$ & 
     $\gamma n \rightarrow \pi^+ \pi^- \pi^0 \pi^0 n$ \\
24 & $\gamma p \rightarrow \pi^+ \pi^+ \pi^- \pi^- p$ & 
     $\gamma n \rightarrow \pi^+ \pi^+ \pi^- \pi^- n$ \\
25 & $\gamma p \rightarrow \pi^+ \pi^0 \pi^0 \pi^0 n$ & 
     $\gamma n \rightarrow \pi^- \pi^0 \pi^0 \pi^0 p$ \\
26 & $\gamma p \rightarrow \pi^+ \pi^+ \pi^- \pi^0 n$ & 
     $\gamma n \rightarrow \pi^+ \pi^- \pi^- \pi^0 p$ \\
27 & $\gamma p \rightarrow \pi^0 \pi^0 \pi^0 \pi^0 \pi^0 p$ & 
     $\gamma n \rightarrow \pi^0 \pi^0 \pi^0 \pi^0 \pi^0 n$ \\
28 & $\gamma p \rightarrow \pi^+ \pi^- \pi^0 \pi^0 \pi^0 p$ & 
     $\gamma n \rightarrow \pi^+ \pi^- \pi^0 \pi^0 \pi^0 n$ \\
29 & $\gamma p \rightarrow \pi^+ \pi^+ \pi^- \pi^- \pi^0 p$ & 
     $\gamma n \rightarrow \pi^+ \pi^+ \pi^- \pi^- \pi^0 n$ \\
30 & $\gamma p \rightarrow \pi^+ \pi^0 \pi^0 \pi^0 \pi^0 n$ & 
     $\gamma n \rightarrow \pi^- \pi^0 \pi^0 \pi^0 \pi^0 p$ \\
31 & $\gamma p \rightarrow \pi^+ \pi^+ \pi^- \pi^0 \pi^0 n$ & 
     $\gamma n \rightarrow \pi^+ \pi^- \pi^- \pi^0 \pi^0 p$ \\
32 & $\gamma p \rightarrow \pi^+ \pi^+ \pi^+ \pi^- \pi^- n$ & 
     $\gamma n \rightarrow \pi^+ \pi^+ \pi^- \pi^- \pi^- p$ \\
33 & $\gamma p \rightarrow \pi^0 \pi^0 \pi^0 \pi^0 \pi^0 \pi^0 p$ & 
     $\gamma n \rightarrow \pi^0 \pi^0 \pi^0 \pi^0 \pi^0 \pi^0 n$ \\
34 & $\gamma p \rightarrow \pi^+ \pi^- \pi^0 \pi^0 \pi^0 \pi^0 p$ & 
     $\gamma n \rightarrow \pi^+ \pi^- \pi^0 \pi^0 \pi^0 \pi^0 n$ \\
35 & $\gamma p \rightarrow \pi^+ \pi^+ \pi^- \pi^- \pi^0 \pi^0 p$ & 
     $\gamma n \rightarrow \pi^+ \pi^+ \pi^- \pi^- \pi^0 \pi^0 n$ \\
36 & $\gamma p \rightarrow \pi^+ \pi^+ \pi^+ \pi^- \pi^- \pi^- p$ & 
     $\gamma n \rightarrow \pi^+ \pi^+ \pi^+ \pi^- \pi^- \pi^- n$ \\
37 & $\gamma p \rightarrow \pi^+ \pi^0 \pi^0 \pi^0 \pi^0 \pi^0 n$ & 
     $\gamma n \rightarrow \pi^- \pi^0 \pi^0 \pi^0 \pi^0 \pi^0 p$ \\
38 & $\gamma p \rightarrow \pi^+ \pi^+ \pi^- \pi^0 \pi^0 \pi^0 n$ & 
     $\gamma n \rightarrow \pi^+ \pi^- \pi^- \pi^0 \pi^0 \pi^0 p$ \\
39 & $\gamma p \rightarrow \pi^+ \pi^+ \pi^+ \pi^- \pi^- \pi^0 n$ & 
     $\gamma n \rightarrow \pi^+ \pi^+ \pi^- \pi^- \pi^- \pi^0 p$ \\
40 & $\gamma p \rightarrow \pi^0 \pi^0 \pi^0 \pi^0 \pi^0 \pi^0 \pi^0 p$ & 
     $\gamma n \rightarrow \pi^0 \pi^0 \pi^0 \pi^0 \pi^0 \pi^0 \pi^0 n$ \\
41 & $\gamma p \rightarrow \pi^+ \pi^- \pi^0 \pi^0 \pi^0 \pi^0 \pi^0 p$ & 
     $\gamma n \rightarrow \pi^+ \pi^- \pi^0 \pi^0 \pi^0 \pi^0 \pi^0 n$ \\
42 & $\gamma p \rightarrow \pi^+ \pi^+ \pi^- \pi^- \pi^0 \pi^0 \pi^0 p$ & 
     $\gamma n \rightarrow \pi^+ \pi^+ \pi^- \pi^- \pi^0 \pi^0 \pi^0 n$ \\
43 & $\gamma p \rightarrow \pi^+ \pi^+ \pi^+ \pi^- \pi^- \pi^- \pi^0 p$ & 
     $\gamma n \rightarrow \pi^+ \pi^+ \pi^+ \pi^- \pi^- \pi^- \pi^0 n$ \\
44 & $\gamma p \rightarrow \pi^+ \pi^0 \pi^0 \pi^0 \pi^0 \pi^0 \pi^0 n$ & 
     $\gamma n \rightarrow \pi^- \pi^0 \pi^0 \pi^0 \pi^0 \pi^0 \pi^0 p$ \\
45 & $\gamma p \rightarrow \pi^+ \pi^+ \pi^- \pi^0 \pi^0 \pi^0 \pi^0 n$ & 
     $\gamma n \rightarrow \pi^+ \pi^- \pi^- \pi^0 \pi^0 \pi^0 \pi^0 p$ \\
46 & $\gamma p \rightarrow \pi^+ \pi^+ \pi^+ \pi^- \pi^- \pi^0 \pi^0 n$ & 
     $\gamma n \rightarrow \pi^+ \pi^+ \pi^- \pi^- \pi^- \pi^0 \pi^0 p$ \\
47 & $\gamma p \rightarrow \pi^+ \pi^+ \pi^+ \pi^+ \pi^- \pi^- \pi^- n$ & 
     $\gamma n \rightarrow \pi^+ \pi^+ \pi^+ \pi^- \pi^- \pi^- \pi^- p$ \\
48 &$\gamma p \rightarrow \pi^0 \pi^0 \pi^0 \pi^0 \pi^0 \pi^0 \pi^0 \pi^0 p$ & 
    $\gamma n \rightarrow \pi^0 \pi^0 \pi^0 \pi^0 \pi^0 \pi^0 \pi^0 \pi^0 n$ \\
49 &$\gamma p \rightarrow \pi^+ \pi^- \pi^0 \pi^0 \pi^0 \pi^0 \pi^0 \pi^0 p$ & 
    $\gamma n \rightarrow \pi^+ \pi^- \pi^0 \pi^0 \pi^0 \pi^0 \pi^0 \pi^0 n$ \\
50 &$\gamma p \rightarrow \pi^+ \pi^+ \pi^- \pi^- \pi^0 \pi^0 \pi^0 \pi^0 p$ & 
    $\gamma n \rightarrow \pi^+ \pi^+ \pi^- \pi^- \pi^0 \pi^0 \pi^0 \pi^0 n$ \\
51 &$\gamma p \rightarrow \pi^+ \pi^+ \pi^+ \pi^- \pi^- \pi^- \pi^0 \pi^0 p$ & 
    $\gamma n \rightarrow \pi^+ \pi^+ \pi^+ \pi^- \pi^- \pi^- \pi^0 \pi^0 n$ \\
52 &$\gamma p \rightarrow \pi^+ \pi^+ \pi^+ \pi^+ \pi^- \pi^- \pi^- \pi^- p$ & 
    $\gamma n \rightarrow \pi^+ \pi^+ \pi^+ \pi^+ \pi^- \pi^- \pi^- \pi^- n$ \\
53 &$\gamma p \rightarrow \pi^+ \pi^0 \pi^0 \pi^0 \pi^0 \pi^0 \pi^0 \pi^0 n$ & 
    $\gamma n \rightarrow \pi^- \pi^0 \pi^0 \pi^0 \pi^0 \pi^0 \pi^0 \pi^0 p$ \\
54 &$\gamma p \rightarrow \pi^+ \pi^+ \pi^- \pi^0 \pi^0 \pi^0 \pi^0 \pi^0 n$ & 
    $\gamma n \rightarrow \pi^+ \pi^- \pi^- \pi^0 \pi^0 \pi^0 \pi^0 \pi^0 p$ \\
55 &$\gamma p \rightarrow \pi^+ \pi^+ \pi^+ \pi^- \pi^- \pi^0 \pi^0 \pi^0 n$ & 
    $\gamma n \rightarrow \pi^+ \pi^+ \pi^- \pi^- \pi^- \pi^0 \pi^0 \pi^0 p$ \\
56 &$\gamma p \rightarrow \pi^+ \pi^+ \pi^+ \pi^+ \pi^- \pi^- \pi^- \pi^0 n$ & 
    $\gamma n \rightarrow \pi^+ \pi^+ \pi^+ \pi^- \pi^- \pi^- \pi^- \pi^0 p$ \\
\hline\hline
\end{tabular}
\end{small}
\end{center}

\clearpage

{\noindent
and described in \cite{GEANT4}) 
to calculate its decay 
instead of considering a 
preequilibrium stage 
followed by 
evaporation from compound nuclei, as described above.
}
LAQGSM03.01 considers also coalescence of complex
particles up to $^4$He from energetic nucleons emitted during the INC,
using an updated coalescence model in comparison with the version
described in \cite{DCM}.

We have developed also ``S" and ``G" versions of LAQGSM03.01, namely:
LAQGSM03.S1 and LAQGSM03.G1 \cite{ResNote06}.
LAQGSM03.S1 \cite{ResNote06} is exactly the same as LAQGSM03.01,
but considers also multifragmentation of excited nuclei produced
after the preequilibrium stage of reactions, when their excitation
energy is above 2A MeV, using the Statistical Multifragmentation
Model (SMM) by Botvina {\it et al.}\ \cite{SMM} (the `S" in the extension
of LAQGSM03.S1 stands for SMM).
LAQGSM03.G1 \cite{ResNote06} is exactly the same as LAQGSM03.01,
but uses the fission-like binary-decay model GEMINI
of Charity {\it et al.}\ \cite{GEMINI},
which considers evaporation of all possible fragments,
instead of using the GEM2 model \cite{GEM2}
(the ``G" stands for GEMINI).

\begin{figure}[th]
\vspace{-0.cm}
\centerline{\hspace{-0mm} \epsfxsize 143mm \epsffile{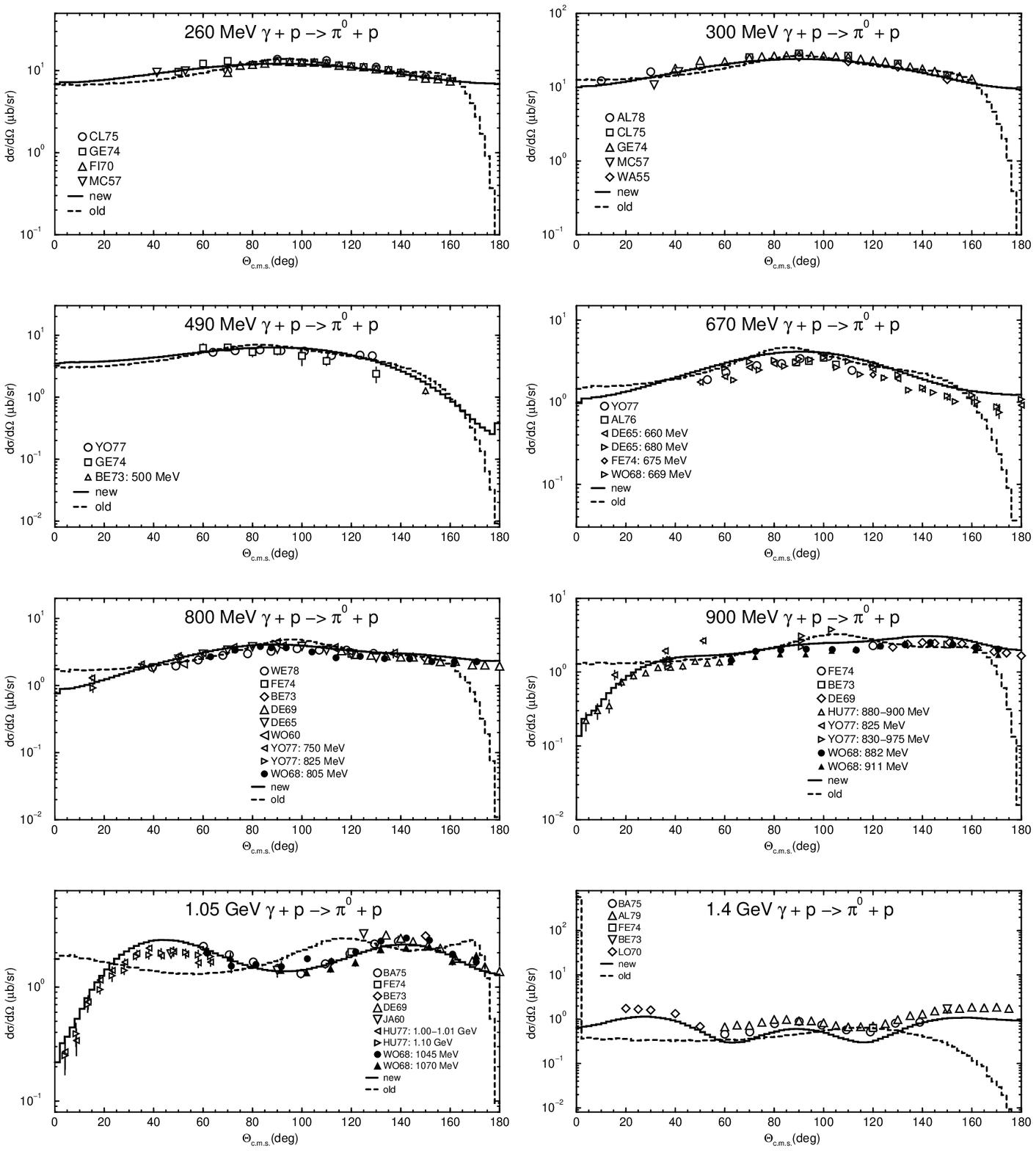}} 
\vspace{-0mm}
{\bf Figure 1.}
Example of eight
angular distributions of $\pi^{0}$ from
$\gamma p \to \pi^{0} p$  
as functions
of $\Theta^{\pi}_{c.m.s}$ at photon energies from 260 MeV to 1.4 GeV.
The dashed lines show the old approximations of the Dubna INC \cite{Dubna69}   
while the solid lines are
our new approximations incorporated into LAQGSM03.01
(and into CEM03 and LAQGSM03 \cite{JNRS05}).
References to
experimental data shown by different symbols
may be found in our recent work \cite{JNRS05}.
\end{figure}

\clearpage
{\noindent \bf \large 3. Particle Spectra}\\

We have tested LAQGSM03.01 and its ``S" and ``G" versions
extended to describe high-energy photonuclear reactions
against practically all experimental
data above 1 GeV we were able to find in the literature, but we
limit ourselves here to presenting only several exemplary results.

Let us start with comparing results by LAQGSM03.01 with experimental
data at relatively low energies, of about 1 GeV, where other photonuclear
models also do work, then move gradually to reactions at higher
energies.
Figure 2 presents examples of proton spectra  
from bremsstrahlung interaction with carbon at $E^{max} = 1050$ MeV.
One can see that LAQGSM03.01 describes as
well as CEM03.01 does the measured proton spectra and agrees with 
the data better than the direct knockout model \cite{Boal81}
and the quasi-deuteron calculations \cite{Matthews66} do.

\begin{figure}[th]
\begin{minipage}{8.0cm}
\vbox to 9.8cm {
\hspace*{-3mm}
\includegraphics[width=75mm,angle=-0]{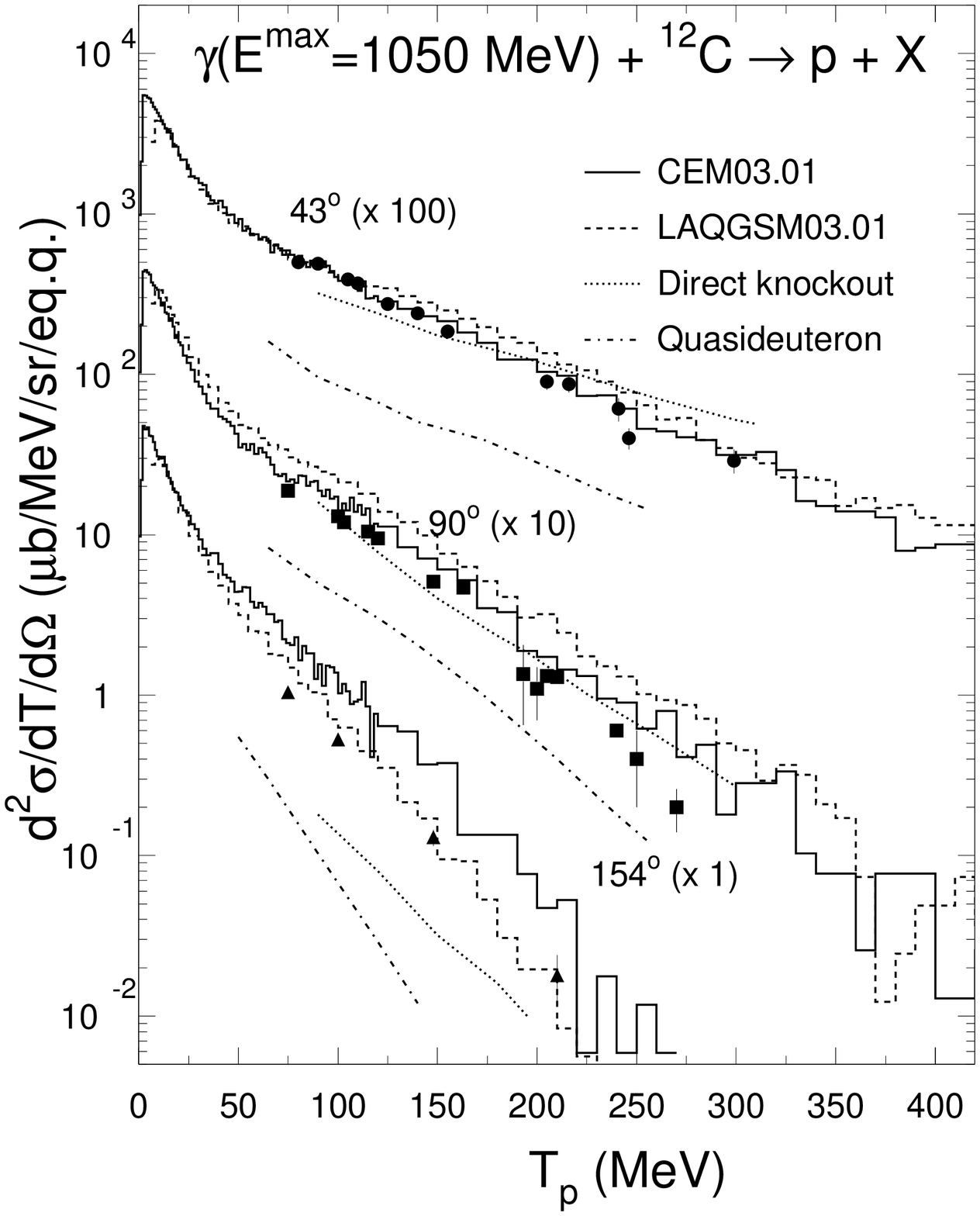}}
\end{minipage}
\hfill
\begin{minipage}{87mm}
\vspace*{-6mm}
\begin{small}
{\bf Figure 2.}
Comparison of measured 
\cite{Olson60}
differential cross section for proton photoproduction on
carbon at 43$^{\circ}$, 90$^{\circ}$, and 154$^{\circ}$ 
by bremsstrahlung photons with $E^{max} = 1.05$ GeV
(symbols) with CEM03.01 (solid histograms)
and LAQGSM03.01 (dashed histograms),
and predictions by the direct knockout model 
\cite{Boal81} 
(dotted lines) and a quasi-deuteron calculation 
\cite{Matthews66} 
(dot-dashed lines), respectively.
The experimental data and results by the direct knockout and
quasi-deuteron models are taken from Fig.\ 5 of Ref.\ 
\cite{Boal81}.\\
\\
\end{small}
\noindent{

Figure 3 presents examples of proton and pion spectra from
photonuclear reactions at higher energies, namely
proton spectra at 60, 90, and 150 degrees from
interaction of bremsstrahlung $\gamma$ quanta of maximum energy $E^{max} =
2.0$, 3.0, and 4.5  GeV with $^{12}$C,  $^{27}$Al,  $^{63}$Cu, and $^{208}$Pb
and pion spectra from
interaction of the same  bremsstrahlung $\gamma$ quanta
with $^{12}$C, $^{63}$Cu, and $^{208}$Pb.
}
These reactions are of a higher interest to us than the one
presented in Fig. 2 for the following reason: The proton and pion
spectra presented in Fig. 3 were measured more than 25 years ago 
at the Yerevan Physics Institute \cite{Alanakyan77}-\cite{Alanakyan80} 
with  a major hope  of revealing some ``exotic" or unknown
\end{minipage}

\vspace*{-1.7mm}
 mechanisms of nuclear reactions leading to
the production of so called ``cumulative"
({\it i.e.}, kinematically forbidden for quasi-free intranuclear 
projectile-nucleon collisions) particles.
 At the time the
measurements  \cite{Alanakyan77}-\cite{Alanakyan80} were done,
there were not high-energy photonuclear reaction models available
in the literature, therefore these data were not analyzed
so far by any models.
We do not know of
any publication or oral presentation where these measurements
were reproduced by a theoretical model, event generator,
or transport code.
It is noteworthy that LAQGSM03.01 describes quite well all 
the spectra measured both in the cumulative and
non-cumulative regions
(as it does, e.g., for particles emitted from interaction of 400 GeV protons
with nuclei; see \cite{Pavia05} for details) 
in a single approach, without any fitting or free parameters, 
and without involving any ``exotic" reaction mechanisms.

\hspace*{4mm}
These results do not imply, of course, that the  $\gamma$- or
proton-nucleus interaction physics
is completely described by the reaction mechanisms considered
by LAQGSM03.01. Our present results do not exclude some
contribution to the production of these particles from other 
reaction mechanisms not considered by LAQGSM03.01.
But the contribution from ``exotic" mechanisms to cumulative particle
production from these high-energy reactions seems to be small; inclusive 
particle spectra are not sensitive enough for an
unambiguous determination of the mechanisms of particle production,
just as observed heretofore at intermediate and low energies \cite{NP94}.
\\
\\
{\noindent \bf \large 4. Product Yields and Recoil Properties}\\
\\
\hspace*{4mm}
This section presents several examples of cross sections
(yields) of the nuclides produced in high-energy photonuclear
reactions and of their recoil properties. 
Fig. 4 shows a comparison of the measured
\vspace*{-5mm}
\end{figure}

\clearpage
\begin{figure}[th]
\vspace*{-10mm}
\centerline{\hspace{-8mm}
\psfig{figure=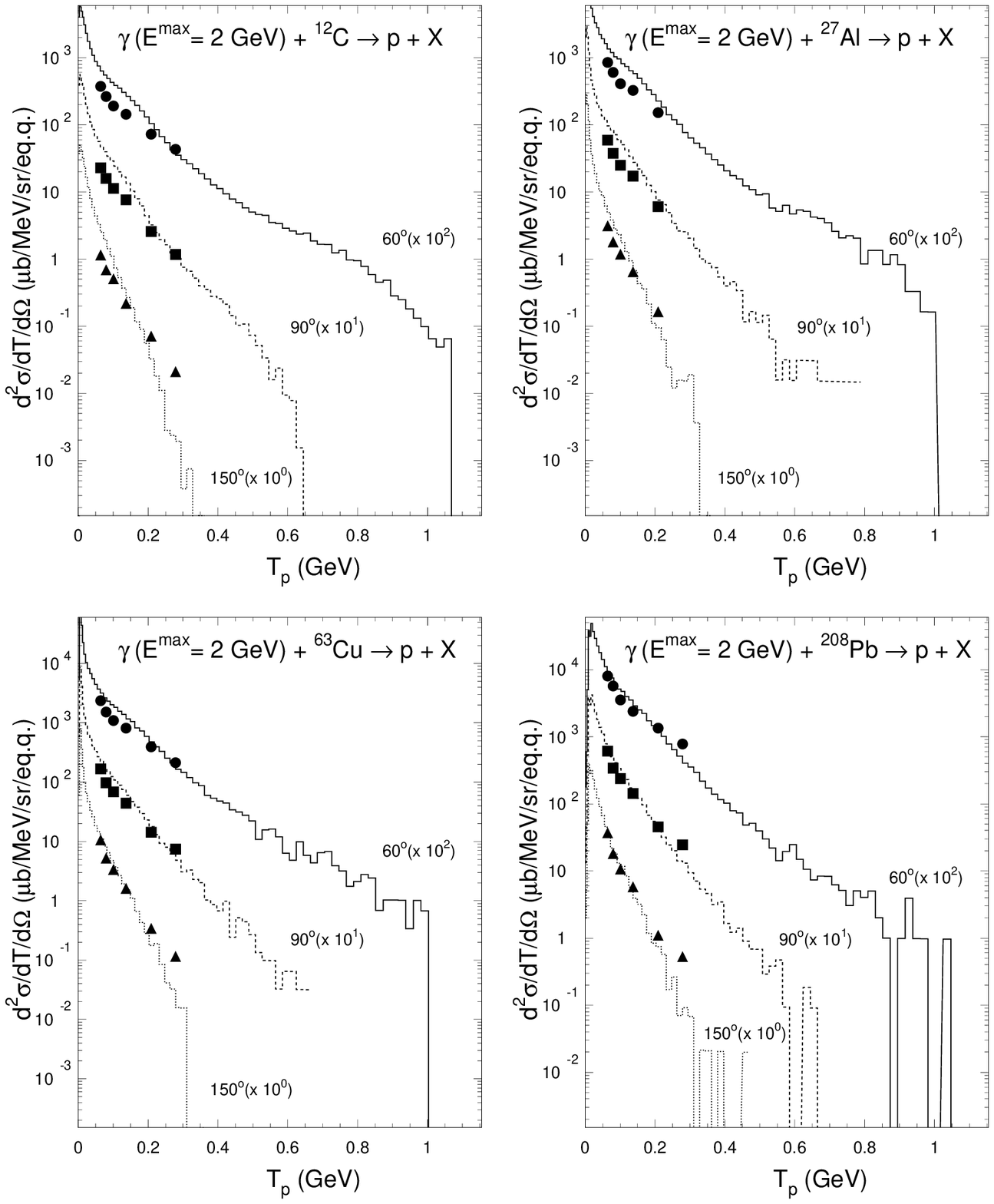,width=92mm,bbllx=42pt,bblly=63pt
,bburx=551pt,bbury=777pt,angle=-0}\hspace{-8mm}
\psfig{figure=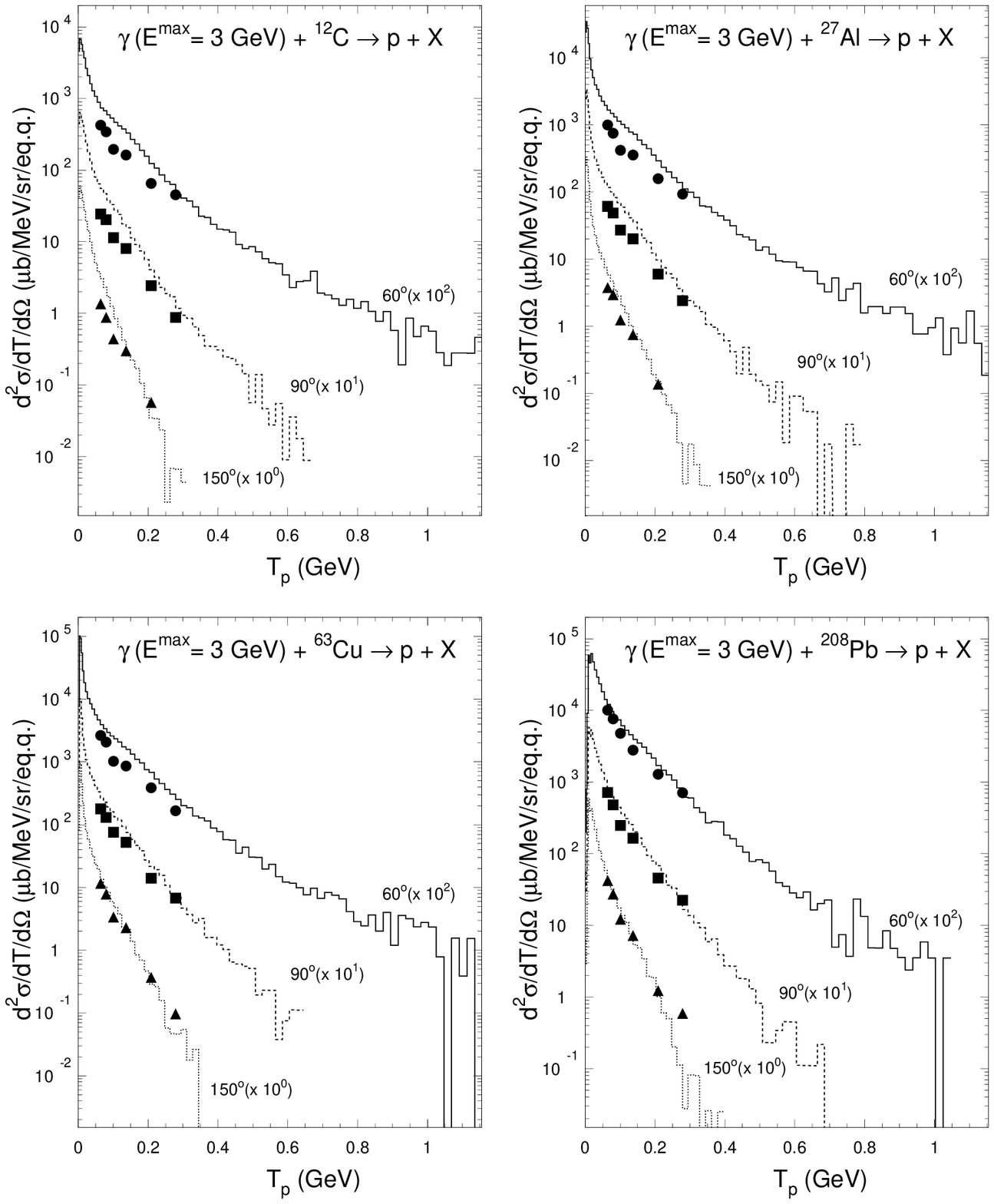,width=92mm,bbllx=42pt,bblly=63pt
,bburx=551pt,bbury=777pt,angle=-0}\hspace{-8mm}}
\vspace{-30mm}
\centerline{\hspace{-8mm}
\psfig{figure=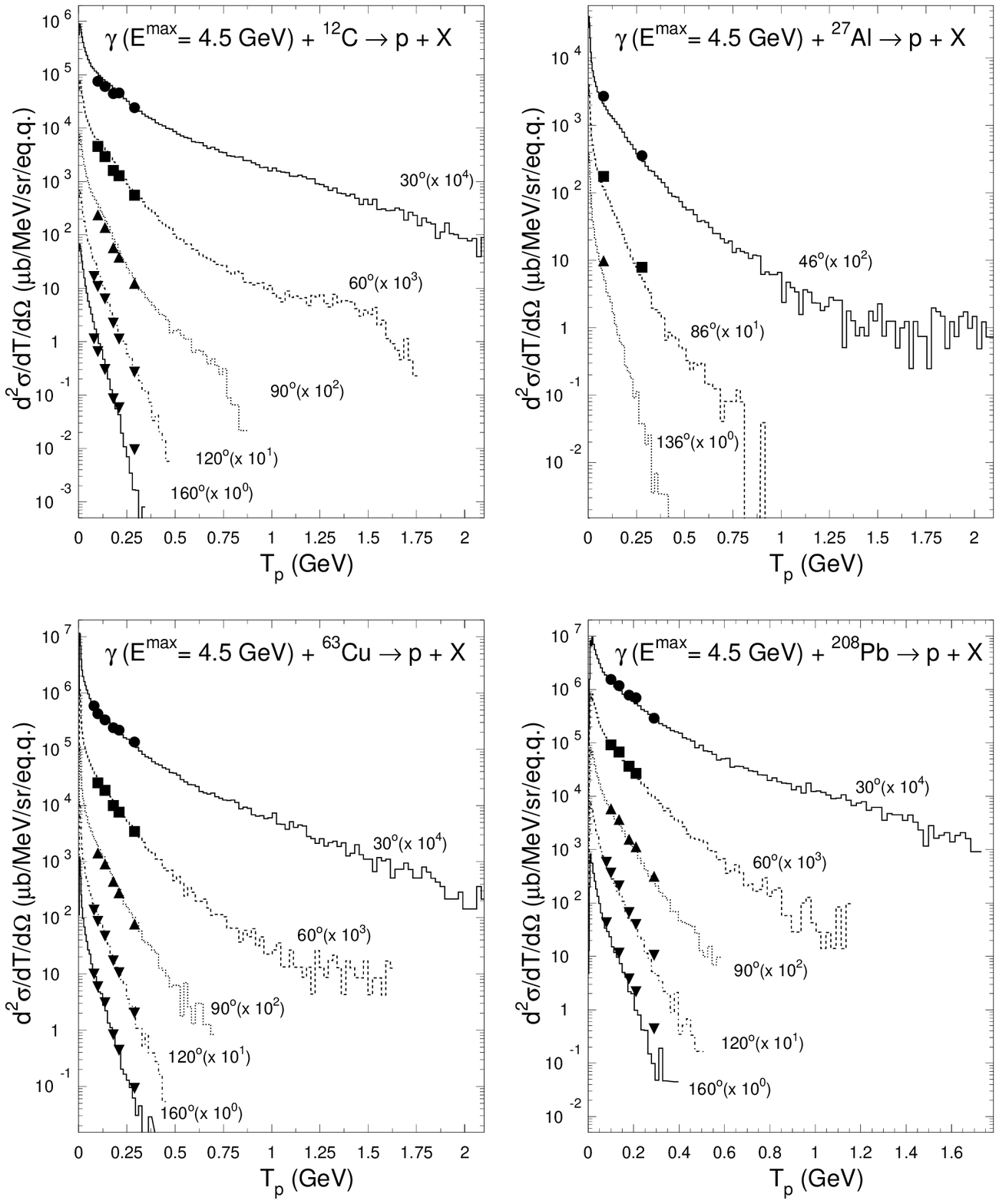,width=92mm,bbllx=42pt,bblly=63pt
,bburx=551pt,bbury=777pt,angle=-0}\hspace{-8mm}
\psfig{figure=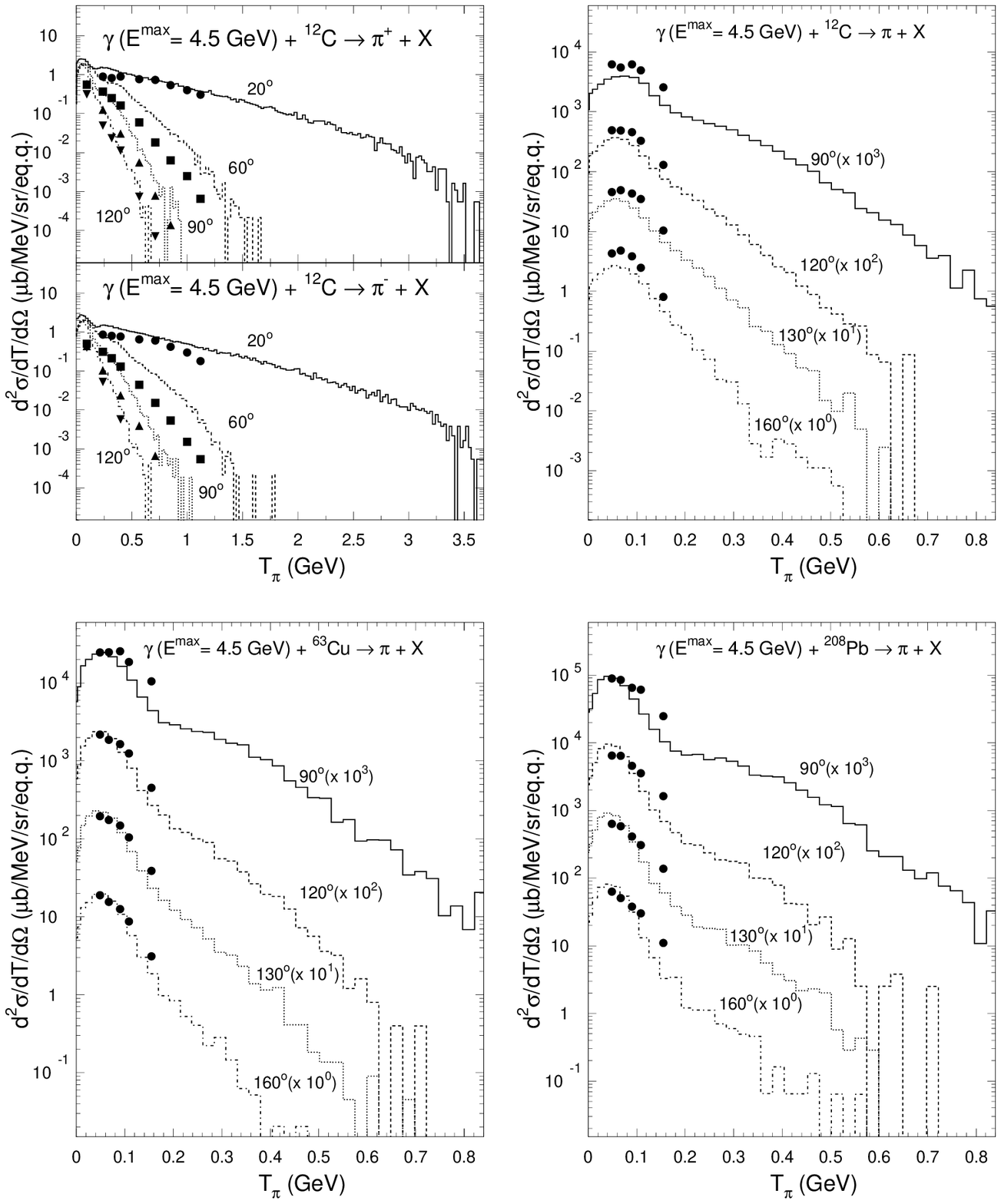,width=92mm,bbllx=42pt,bblly=63pt
,bburx=551pt,bbury=777pt,angle=-0}\hspace{-8mm}}

\vspace*{-26mm}
{\bf Figure 3.}
Proton spectra at 60, 90, and 150 degrees from
interaction of bremsstrahlung $\gamma$ quanta of maximum energy $E^{max} =
2.0$, 3.0, and 4.5  GeV with $^{12}$C,  $^{27}$Al,  $^{63}$Cu, and $^{208}$Pb
(left and right top panels and left bottom panel of four plots).
Spectra of  $\pi^+$ and $\pi^-$
produced by $E^{max} = 4.5$ GeV bremsstrahlung 
on $^{12}$C 
and spectra of charged pions (both $\pi^+$ and $\pi^-$)
from
interaction of the same  bremsstrahlung $\gamma$ quanta
with $^{12}$C, $^{63}$Cu, and $^{208}$Pb
(right bottom panel of four plots).
Experimental values shown by symbols are from 
\cite{Alanakyan77}-\cite{Alanakyan80}
while histograms show results by LAQGSM03.01.
To the best of our knowledge, we are able to describe
these data with LAQGSM03.01 for the first time (see text).
\end{figure}

\newpage
\begin{figure}[h]
\begin{minipage}{11.0cm}
\vbox to 79mm {
\vspace*{-20mm}
\hspace*{-5mm}
\includegraphics[width=110mm,angle=-0]{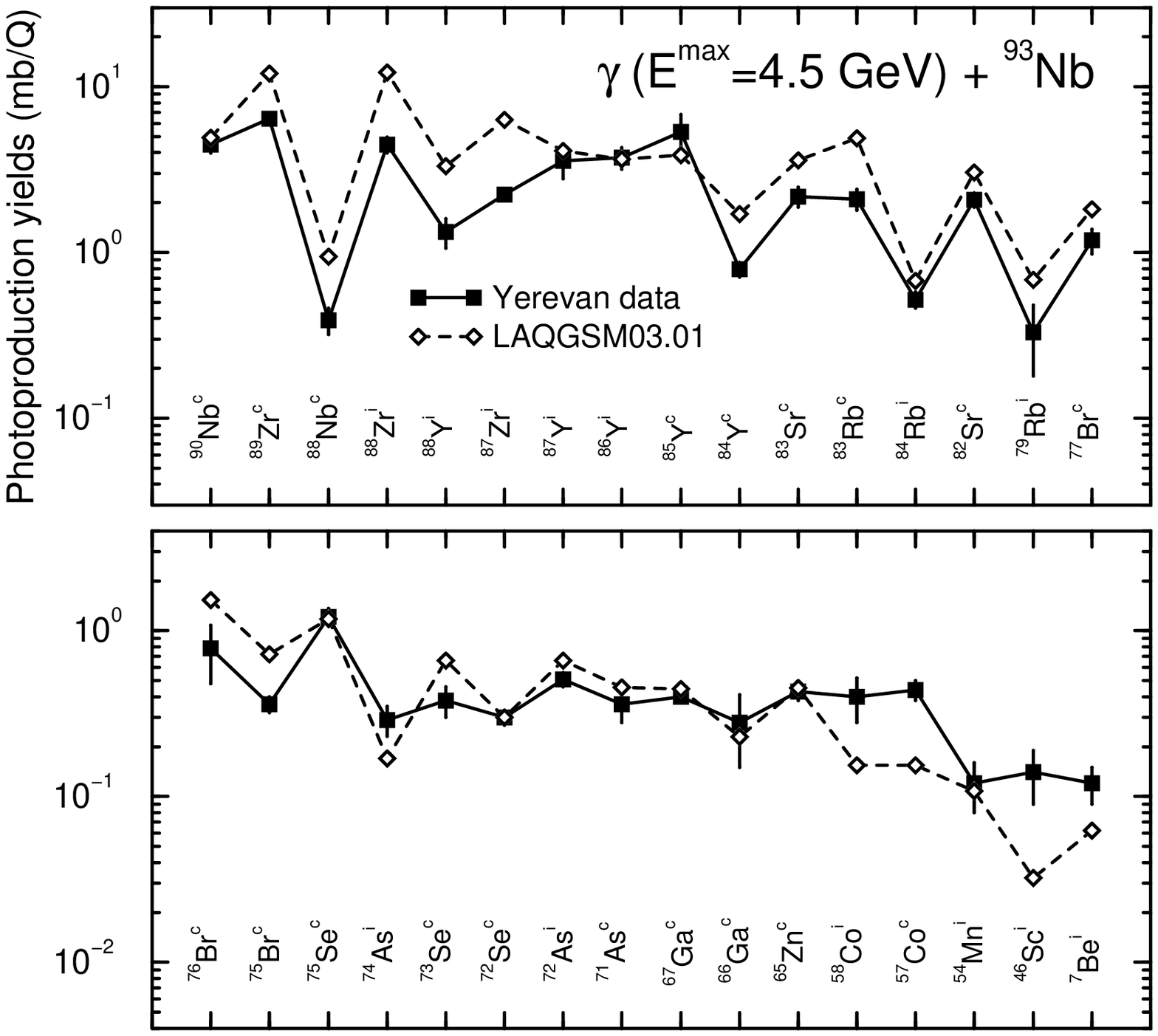}}
\end{minipage}
\hfill
\begin{minipage}{6.0cm}
\vspace*{-13mm}
\begin{small}
{\bf Figure 4.}
Detailed comparison between experimental yields 
\cite{Vartapetyan81}
and those calculated by 
LAQGSM03.01 of radioactive products from the interaction
of bremsstrahlung $\gamma$ quanta of maximum energy 4.5 GeV with 
$^{93}$Nb. The cumulative yields are labeled as ``c" while
the independent cross sections, as ``i".
To the best of our knowledge, using LAQGSM03.01 we are able 
to describe these data for the first time.\\
\\
\end{small}
\noindent{
 production cross sections
from interaction
of bremsstrahlung $\gamma$ quanta of maximum energy 4.5 GeV with 
$^{93}$Nb
\cite{Vartapetyan81}     
with results by LAQGS03.01.
We see that LAQGSM03.01 describes most of the measured yields
quite well, with only several cases when the calculations
differ from the data within a factor of two to three, that is also not too 
bad,
considering that all calculations are in the absolute value, without any
fitting of parameters or normalization.
}

\end{minipage}
\vspace*{0.5mm}
\hspace*{4mm}
Fig. 5 shows a similar reaction, induced by $E^{max} = 4.5$ GeV
bremsstrahlung on $^{65}$Cu, but measured regarding the recoil properties
of the produced nuclides \cite{Arakelyan91}. No product yields were
measured from this reaction, so the top-left plot in Fig. 5 shows only
predictions by LAQGSM03.01 and its ``S" and ``G" versions for this
quantity. The left plot in the middle row of Fig. 5 shows
predictions by LAQGSM03.01 and its ``S" and ``G" versions for
the mean laboratory angle of products. We see quite a big
difference for this characteristics
between predictions by versions of the model with
different treatment of the evaporation stage of reaction (compare
the ``standard", LAQGSM03.01, results that uses GEM2
with results by its ``G" version that uses GEMINI)
and between results calculated without (LAQGSM0.01) and with taking
into account multifragmentation of highly-excited compound nuclei
(LAQGSM03.S1). Unfortunately, this quantity was not measured and
we can not uncover the ``real" reaction mechanisms based on these
results until experimental data are available.

\hspace{4mm}
The other three plots in Fig. 5 show results 
by LAQGSM03.01 and its ``G" and ``S" versions for the 
Z-averaged A-dependence of the
mean laboratory velocity $v_z$ of products, 
the R=F/B ratio of the forward product cross sections to the backward ones,
and their mean laboratory kinetic energy
compared with available data \cite{Arakelyan91}. We see a very good
agreement between calculations and the data for the mean kinetic energy,
a not so good agreement for the mean parallel velocity $v_z$ of products,
and a poor agreement for R=F/B. Let us note that in the experimental work
\cite{Arakelyan91}, these characteristics were measured only for
several final isotopes, while results of calculations shown in Fig. 5
are for the Z-averaged A-dependence of all possible products.
If to compare the data with calculations for only the measured
products, isotope-by-isotope, the agreement between measurements and
calculations becomes better. However, 
as can be seen from Fig.\ 6, there is still some disagreement between
experimental data and theoretical results by all three
versions of LAQGSM considered here for the values of R=F/B,
similar to results obtained for reactions induced
by protons and deuterons \cite{NPA06}.
In making such comparisons, we first recognize
that the experiment and the calculations differ in that:
1)
the experimental data were extracted assuming
the ``two-step vector model" (see references and details in 
\cite{Arakelyan91,NPA06}), 
while the LAQGSM calculations were done
without the assumptions of this model;
2) the measurements were performed on foils (thick targets),
while the calculations were done for interactions
of photons (protons and deuterons, in the case of Ref. \cite{NPA06})
with nuclei (thin targets).
These differences must be considered before assessing
possible deficiencies in the models.

\end{figure}

\newpage
\begin{figure}[h]
\begin{minipage}{13.0cm}
\vspace*{-8mm}
\vbox to 160mm {
\hspace*{-5mm}
\includegraphics[width=130mm,angle=-0]{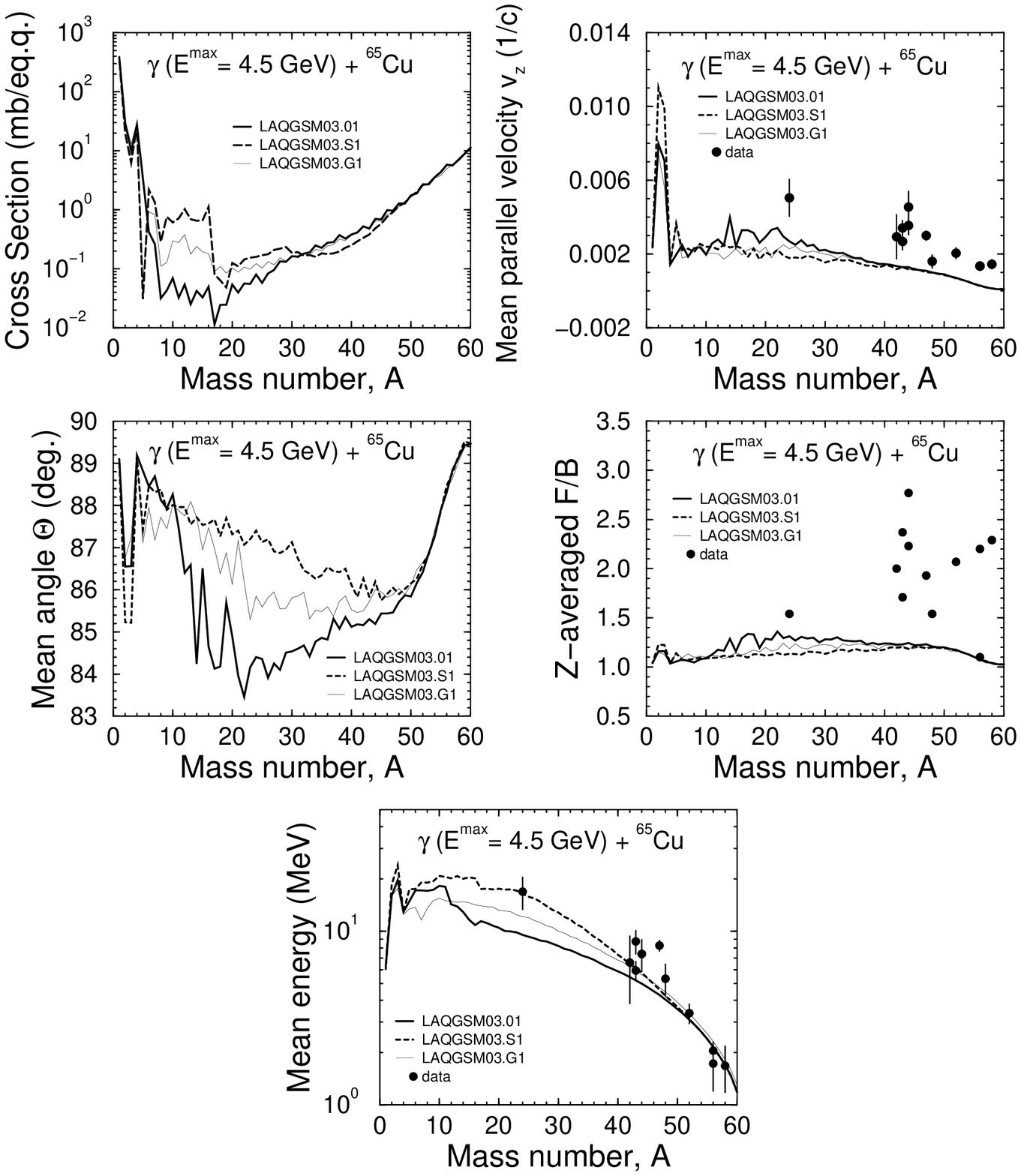}}
\end{minipage}
\hfill
\begin{minipage}{3.9cm}
\vspace*{-15mm}
\begin{small}
{\bf Figure 5.}
Results by LAQGSM03.01 and its ``S" and ``G" versions
for the Z-averaged A-dependence of the 
 mass yield of all products, their  mean parallel
laboratory velocity $v_z$ 
in the beam direction, their mean laboratory angle $\Theta$,  
the R=F/B ratio of the forward product cross sections to the backward ones,
and their mean laboratory kinetic energy for the reaction of
$E^{max} = 4.5$ GeV  bremsstrahlung photons on $^{65}$Cu compared
with available experimental data by Arakelyan {\it et al.}
\cite{Arakelyan91}.\\
\\
\end{small}
\hspace*{4mm}
Finally, Fig. 7 shows a comparison of the calculated
by LAQGSM03.01
mass distribution of products with experimental data
for the highest energy we were able to find data in the
literature, namely for the reaction $E^{max} = 6$ GeV bremsstrahlung gammas
on $^{238}$U \cite{Andersson72,Schroder72}. We see a good agreement 
between the calculations and these high-energy photonuclear reaction
data.

\end{minipage}

\hspace{4mm}
To conclude, let us note that to the best of our knowledge,
we were able to describe with LAQGSM03.01 all the reactions
discussed in this section for the first time:
We do not know of
any publication or oral presentation where these experimental data
were reproduced by a theoretical model, event generator,
or transport code.\\
\\

{\noindent \bf \large 5. Summary} \\
\\
The latest modification of the Los Alamos Quark-Gluon String Model
released in the code LAQGSM03.01 and its ``S" and ``G" versions
were extended to describe photonuclear reactions at energies up
to tens of GeV. We have tested our high-energy photonuclear models
against practically all experimental data above 1 GeV we were able to find 
in the literature and can conclude that they describe the
measured data reasonably well, without re-fitting any parameters.
Our models are not yet ready to describe properly
reactions induced by gammas
of tens of MeV, in the GDR region (though they provide 
not completely wrong results even at such low energies; 
see, e.g., \cite{JNRS05}). 
We are developing now approximations for the GDR region, to extend the
use of our codes for low-energy photonuclear reactions as well.
LAQGSM03.01 was
incorporated recently into the transport code MARS \cite{MARS} and 
is currently
being incorporated into MCNP6 \cite{MCNP6} and MCNPX \cite{MCNPX}.
This would allow others to use LAQGSM03.01 as an event-generator
in these transport codes to simulate high-energy photonuclear
reactions with targets of practically arbitrary geometry and 
nuclide composition.
\end{figure}

\newpage
\begin{figure}[t]
\begin{minipage}{80mm}
\vbox to 80mm {
\vspace*{-00mm}
\hspace*{-5mm}
\includegraphics[width=80mm,angle=-0]{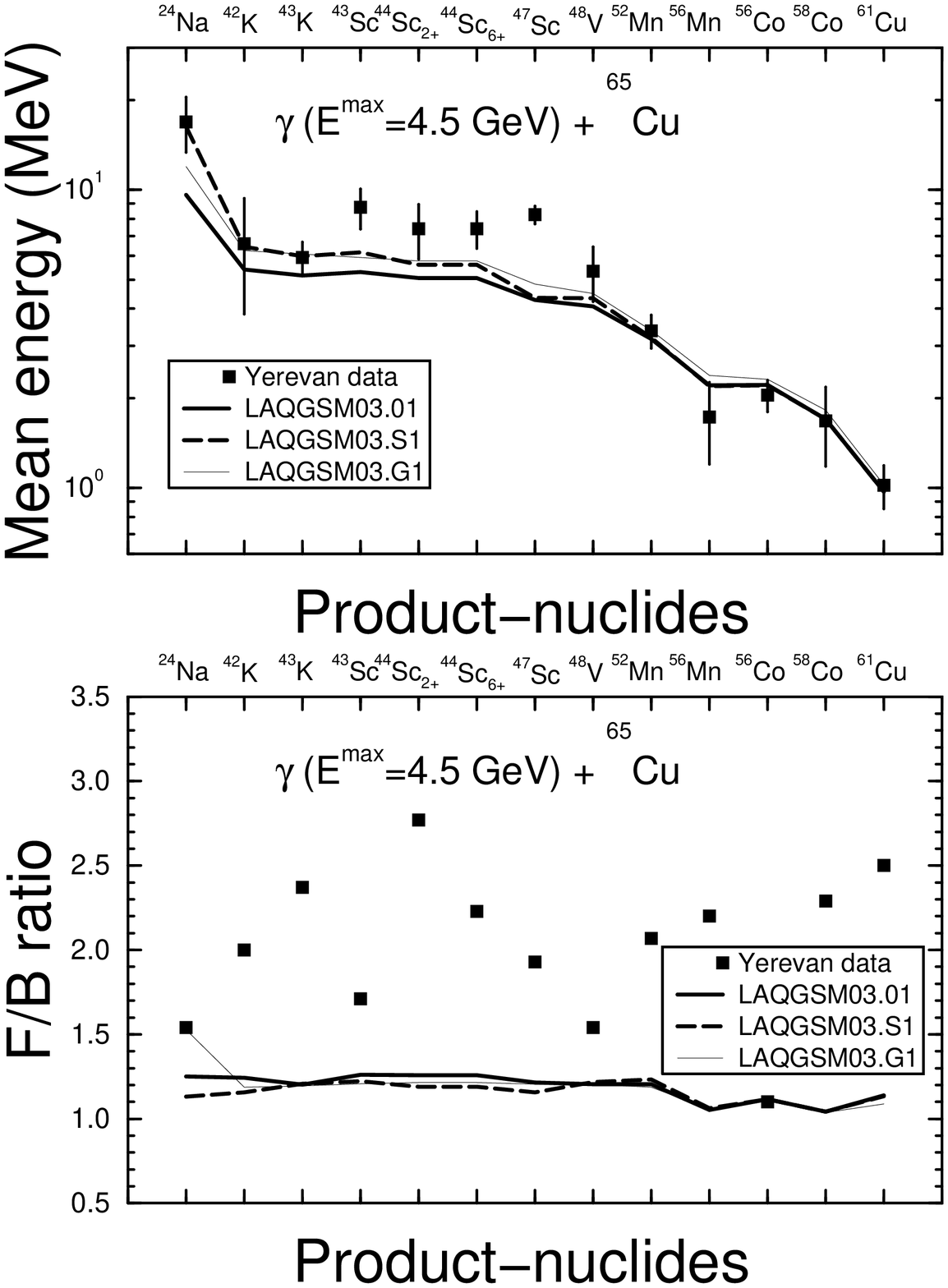}}
\end{minipage}
\hfill
\begin{minipage}{90mm}
\begin{small}
\vspace*{-40mm}
{\bf Figure 6.}
The measured data \cite{Arakelyan91} for 
the mean laboratory kinetic energy and 
the R=F/B ratio of the forward product cross sections to the backward ones
for the same reaction as shown in Fig. 5,
but compared with results by LAQGSM03.01 
and its ``S" and ``G" versions
for only the measured
products (listed on the top of both plots), isotope-by-isotope.\\
\\
\end{small}
\end{minipage}

\begin{minipage}{9.0cm}
\vbox to 9.8cm {
\vspace*{50mm}
\hspace*{-5mm}
\includegraphics[width=70mm,angle=-90]{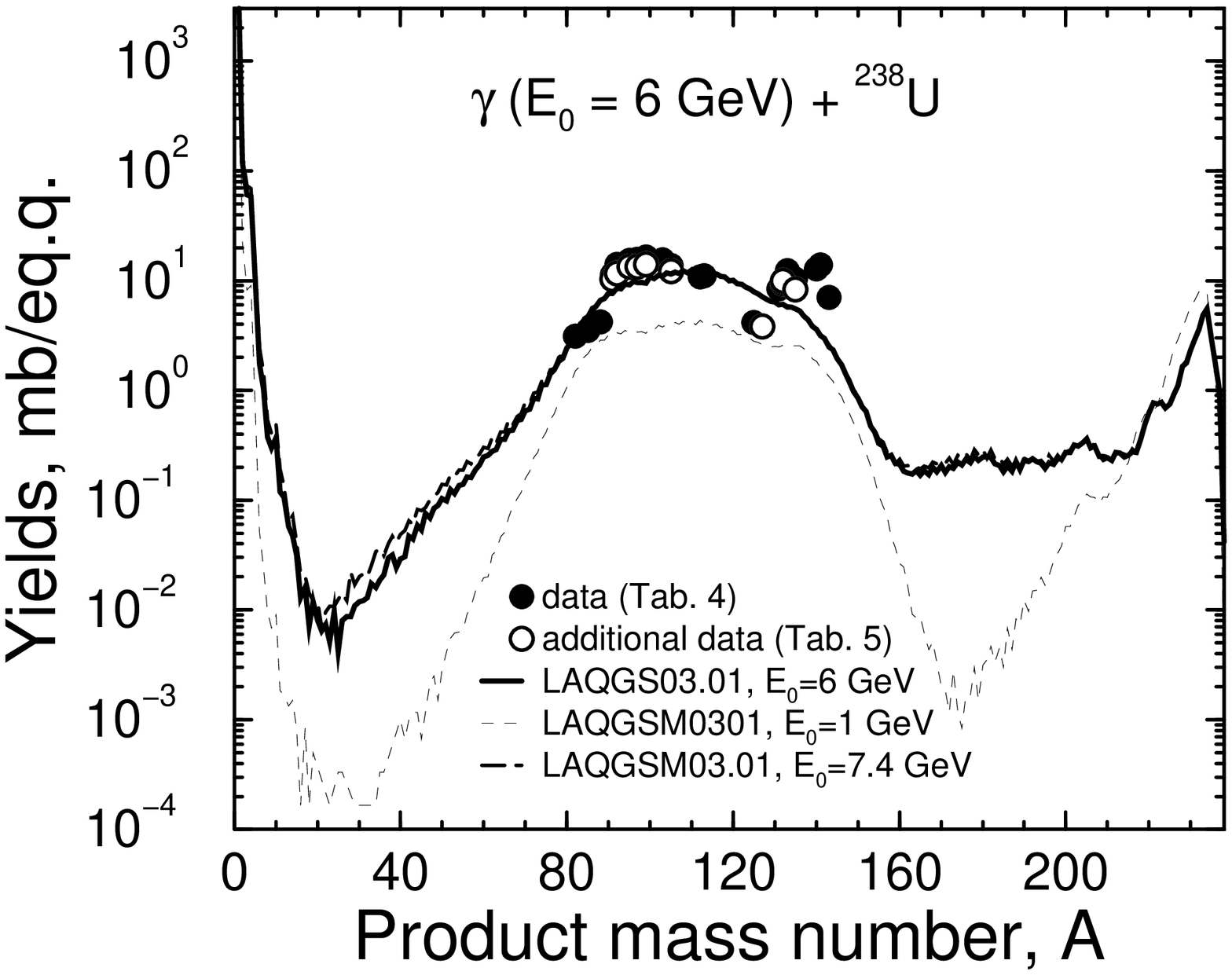}}
\end{minipage}
\hfill
\begin{minipage}{8.0cm}
\vspace*{-20mm}
\begin{small}
{\bf Figure 7.}
Comparison of measured (filled circles) \cite{Andersson72,Schroder72}
mass distribution of the nuclides produced by $E^{max} = 6$ GeV
bremsstrahlung photons on $^{238}$U with results
by LAQGSM03.01 (thick solid line).
Open circles show some additional experimental data measured at other
energies and
compiled from the literature 
in Tab. 5 of Ref. \cite{Andersson72}; 
results by LAQGSM03.01 at  $E^{max} = 1.0$ to 7.4 GeV are shown
for comparison by thin and thick dashed lines, respectively.\\
\\
\end{small}
\end{minipage}
\end{figure}

\begin{figure}
\begin{center}
{\it Acknowledgment}
\end{center}

\noindent
We are grateful to
Dr.\ Igor Pshenichnov for sending us
the $\gamma p$
and $\gamma n$ event generators from their Moscow photonuclear
reaction INC \cite{Iljinov97} and thank our collaborators
Drs. Arnie Sierk, Richard Prael, Nikolai Mokhov, and Mircea Baznat
for useful discussions.
This work was supported by the 
U.\ S.\ Department of Energy and by the
Moldovan-U.\ S.\ Bilateral Grants Program, CRDF Project MP2-3045.
SGM acknowledge partial support from a NASA
Astrophysics Theory Program grant.



\end{figure}
\end{document}